\documentstyle[prl,epsfig,aps,floats]{revtex}

\begin{document}
\draft
\def\overlay#1#2{\setbox0=\hbox{#1}\setbox1=\hbox to \wd0{\hss #2\hss}#1
-2\wd0\copy1}
\twocolumn[\hsize\textwidth\columnwidth\hsize\csname@twocolumnfalse\endcsname

\title{Stabilization of Deterministically Chaotic Systems by Interference and
       Quantum Measurements: The Ikeda Map Case}

\author{Mauro Fortunato\cite{mau}$^{(a)}$ \and Gershon Kurizki$^{(b)}$ \and
        and Wolfgang P. Schleich$^{(a)}$}
\address{(a) Abteilung f\"ur Quantenphysik, Universit\"at Ulm,
          Albert-Einstein-Allee 11, D-89069 Ulm, Germany}
\address{(b) Department of Chemical Physics, The Weizmann Institute of
             Science, Rehovot 76100, Israel}
\date{\today}
\maketitle
\begin{abstract}
We propose a method which can effectively stabilize fixed points in
the classical and quantum dynamics of a phase-sensitive chaotic system
with feedback. It is based on feeding back a selected quantum
sub-ensemble whose phase and amplitude stabilize the otherwise chaotic
dynamics. Although the method is rather general, we apply it to
realizations of the inherently chaotic Ikeda map. One suggested
realization involves the Mach-Zender interferometer with Kerr
nonlinearity. Another realization involves a trapped ion interacting
with laser fields.
\end{abstract}

\vskip2pc]

Deterministic (Hamiltonian) chaos in a classical system manifests
itself by the extreme (exponential) sensitivity of the evolution to
initial conditions, thus making long-time predictions on the dynamical
observables of the system practically impossible~\cite{kn:chaos}.
Yet, it has been recently recognized~\cite{kn:ccw,kn:ccs} that it is possible
to control chaos: by applying specifically designed time-dependent
perturbations, regular periodic orbits can be stabilized within
chaotic phase-space domains~\cite{kn:ccw,kn:ccs}. The quantum dynamics
underlying classical chaos has been extensively investigated for some
time now~\cite{kn:milo}. Of particular relevance here is the finding
\cite{kn:zur} that quantum spread can cause a quasiclassical state
initially confined within regular phase-space regions to eventually
access chaotic regions. It is therefore natural to ask: what is the
quantum analog of classical chaos control and how can we deal with
chaotic behavior induced by quantum spread?

It is our purpose here to propose the quantum control of
deterministically chaotic two-dimensional systems, whose evolution is
governed by their phase ($mod \; 2\pi$), and involves phase-dependent
feedback. The proposed control is based on two elements. First,
creation of phase correlations between the stabilized system and a
similar system acting as a ``stabilizer''. If the stabilized and
stabilizer systems are classically correlated and are not entangled,
then their phase correlations allow us to enforce regular evolution
(stabilization) of an inherently-chaotic stabilized system, by feeding
it with the output of the stabilizer system. However, whenever these
systems are appreciably entangled, this stabilization may not
suffice. If we merely ignore (trace out) the stabilizer states, and
feed back the resulting statistical mixture of stabilized-system
states, then stabilization may eventually fail, because one of the
states in the mixture may venture into a chaotic region of phase space
(see below). This calls for the second element of proposed control: a
quantum measurement of the entangled stabilized and stabilizer
systems, which selects a particular state of the stabilized system and
feeds it back into this system. The chaotic trend due to
entanglement-induced spread is thus suppressed by suitable projections
which localize the system in regular phase-space regions. Such
manipulation of the evolution, i.e., selection of a certain state
corresponding to the desired outcome of the measurement, is the
essence of the conditional measurement (CM) approach to quantum state
control \cite{kn:cm}. The price one has to pay for guiding the
evolution by CMs is that one must perform as many trials as implied by
the success probability of the required CM until it is accomplished
(see below). However, we stress that this price is unavoidable, since
there is no other remedy for the chaotic effects of entanglement.
Since the quantum version of our scheme is non-unitary,
it does not correspond to traditional control theory, in which a
force driving the system can be identified. This essential difference stems
from our objective, namely, to combat the unwanted effects of entanglement.
We finally note that all of the above considerations apply on time-scales
much shorter than dissipation (decoherence) time.

Although the outlined proposal is quite general, we shall apply it
here to systems governed by the Ikeda map~\cite{kn:ike,kn:mol}. To
this end, we shall analyze the Ikeda map quantum-mechanically,
focusing on the quasiclassical limit, thereby extending its existing
classical analysis. Two different realizations will be considered: (i)
a Mach-Zender interferometer with Kerr nonlinearity; (ii) a trapped
ion interacting with traveling- and standing-wave fields.

\begin{figure}
\centerline{\hbox{\epsfig{file=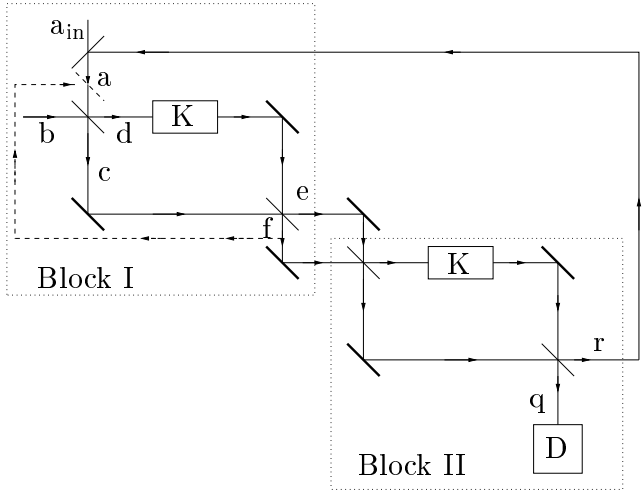,width=3.4in}}}
\vspace{0.1cm}
\caption{Schematic representation of the stabilized (Block I) and
         stabilizer (Block II) systems: two coupled Mach-Zender
         interferometers with Kerr nonlinearities (indicated by K) and
         feedback. The input ports are $a$ and $b$. The $45^\circ$
         thin lines are 50--50 beam splitters, while the $45^\circ$
         thick lines are perfectly reflecting mirrors. D denotes a
         homodyne detector. In the case of the single MZI the output
         $f$ is fed back into port $a$ (via the dashed-line beam
         splitter), while in the case of the total (system +
         stabilizer) setup the output $r$ is fed back into $a$ (via
         the thin-line beam splitter).}
\label{fg:scheme}
\end{figure}

We first consider the following realization of the Ikeda map, which
differs from its standard realization in a ring
cavity~\cite{kn:ike}. The basic block of the proposed realization is a
Mach-Zender interferometer (MZI) with a Kerr-nonlinear element in one
arm, such that the output of one of the ports of this interferometer
is fed back into the input port, along with a fixed input field (see
block I in Fig.~\ref{fg:scheme}). The input-output operator
transformation for this MZI is
\begin{equation}
\left(\begin{array}{c} \hat{f} \\ \hat{e} \end{array} \right)
=ie^{i\hat{\theta}_{\rm I}/2}\left(\begin{array}{cc}
\cos\frac{\hat{\theta}_{\rm I}}{2} & \sin\frac{\hat{\theta}_{\rm
I}}{2} \\ -\sin\frac{\hat{\theta}_{\rm I}}{2} &
\cos\frac{\hat{\theta}_{\rm I}}{2} \end{array} \right) \left(
\begin{array}{c} \hat{a} \\ \hat{b} \end{array} \right)\;,
\label{eq:inout}
\end{equation}
where $\hat{\theta}_{\rm I}=\varphi_{\rm I}+\kappa_{\rm
I}(\hat{b}+i\hat{a})^{\dag} (\hat{b}+i\hat{a})$ is a phase-shift
operator, with a {\it c}-number (refractive) shift $\varphi_{\rm
I}=\chi^0_{\rm I}l_{\rm I}$ and a Kerr shift which is proportional to
the Kerr coefficient $\kappa_{\rm I}=\chi^{(3)}_{\rm I}l_{\rm I}/2$
and to the field intensity $\hat{d}^{\dag}\hat{d}$~\cite{kn:miwa}.

The classical counterpart of this transformation, obtained by
replacing all operators by {\it c}-numbers and setting $b=0$ (vacuum
input in port $b$) can be used to describe repeated feedback of the
output into the input port $a$.  If we choose to feed back the output
$f$ into port $a$ together with fixed external input
(Fig.\ref{fg:scheme}), we obtain the following iterative map
\begin{equation}
a_{j+1}=\frac{1}{\sqrt{2}} \left[ a_{\rm in}+ie^{i \theta_{{\rm I}
j}/2} \cos(\theta_{{\rm I} j}/2){a}_j \right]\;,
\label{eq:itemap}
\end{equation}
where $j=0,1,2,\ldots$ is the iteration number, ${a}_{\rm in}$
represents the fixed input, and the $1/\sqrt{2}$ factor is imposed by
the 50--50 beam splitter at port $a$.

\begin{figure}
\centerline{\hbox{\epsfig{file=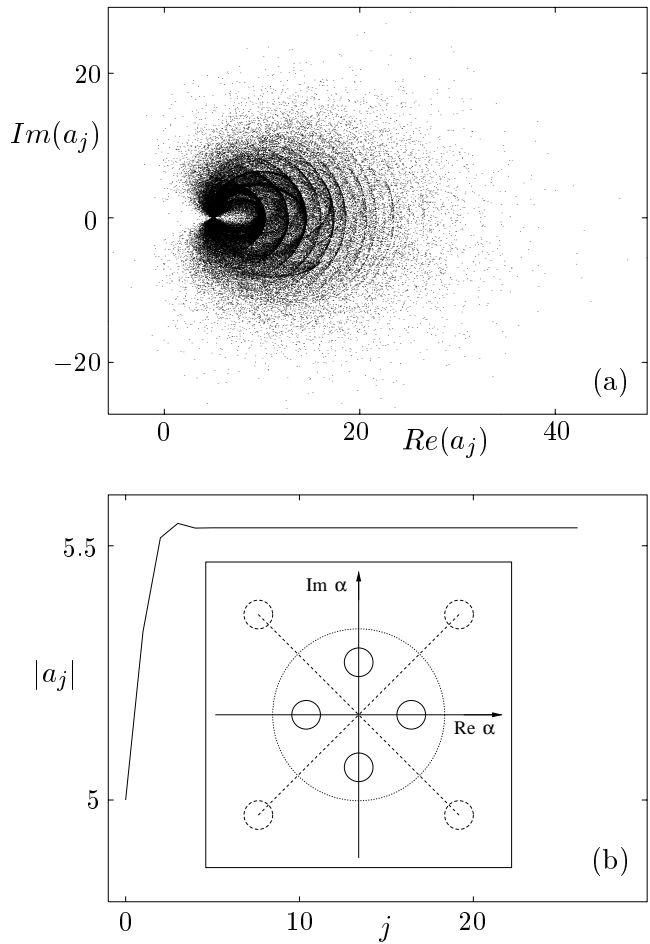,width=3.4in}}}
\vspace{0.1cm}
\caption{(a) Chaotic dynamics of the first block. The ``phase-space''
         picture [$Re(a_j)$--$Im(a_j)$] is plotted for $10^5$
         iterations. The parameters used are
         $\varphi=\pi/2+\varphi_{\rm I}$ with $\varphi_{\rm I}=0.4$,
         $a_{\rm in}=5.0$ and $\kappa=0.1$.  (b) Regular (stabilized)
         dynamics of the stabilized system enforced by the stabilizer
         system. Here the modulus of $a_j$ is plotted versus $j$ for
         the same values of $a_{\rm in}$ and $\kappa$ as in (a), but
         with $\varphi=\pi+\left(\varphi_{\rm I}+\varphi_{\rm
         II}\right)/2 =\pi/2+\varphi_{\rm I}+\delta$ and $R=0.1$. The
         fixed point $(5.53,0.15)$ is reached after few
         iterations. Inset: Phase-plane distribution of a two-mode
         entangled superposition of coherent states for a fractional
         revival. Solid circles are ``Schr\"odinger kittens'' in one
         of the modes, with mean amplitude $\alpha \cos \Phi$, dashed
         circles are their counterparts in the other mode, with mean
         amplitude $\alpha \sin \Phi$. The dotted circle demarcates
         the boundary between the chaotic-domain and the
         regular-domain amplitudes.}
\label{fg:dyn}
\end{figure}

This is a generalization of the Ikeda map in the limit of small Kerr
shifts~\cite{kn:mol}, $\kappa|a_j|^2\ll \varphi$,
\begin{equation}
a_{j+1}=\tilde{a}_{\rm in}+e^{i(\varphi+\kappa |a_j|^2/2)}Ra_j\;,
\label{eq:ikeda}
\end{equation}
with $\tilde{a}_{\rm in}=a_{\rm in}/\sqrt{2}$, in that the feedback
parameter $R$ is now replaced by $\cos(\theta_{{\rm I}
j}/2)/\sqrt{2}$, i.e., it varies with $j$. Its dynamics becomes
chaotic whenever the {\it c}-number parameter $\tilde{a}_{\rm
in}/[1-|\cos(\theta_{{\rm I} j}/2)|]$ is large enough~\cite{kn:ike},
corresponding to excessive feedback, and $\varphi,\kappa \neq 0$ [see
Fig.~\ref{fg:dyn}(a)].

We intend to show how the chaotic dynamics of this device can be made
regular (stabilized) by connecting it to a similar block (block II in
Fig.~\ref{fg:scheme}) via an interface. The second nonlinear MZI
(block II) will then be shown to act as a ``stabilizer''. The
corresponding output-input operator transformations for the second
block are
\begin{equation}
\left(\begin{array}{c} \hat{r} \\ \hat{q} \end{array} \right)
=ie^{i\hat{\theta}_{II}/2}\left(\begin{array}{cc} \cos\hat{\theta}_{II}/2 &
-\sin\hat{\theta}_{II}/2 \\ \sin\hat{\theta}_{II}/2 & \cos\hat{\theta}_{II}/2
\end{array}
\right) \left(\begin{array}{c} \hat{f} \\ \hat{e} \end{array} \right)\;,
\label{eq:outin}
\end{equation}
where
$\hat{\theta}_{\rm II}=\varphi_{\rm II}+\kappa_{\rm II}
(\hat{f}+i\hat{e})^{\dag} (\hat{f}+i\hat{e})$.

In order to achieve stabilization we impose the conditions
\begin{mathletters}
\begin{eqnarray}
\frac{1}{2}\left(\varphi_{\rm I}-\varphi_{\rm II}\right)
&=&\frac{1}{2}\left(\chi^0_{\rm I}l_{\rm I}
-\chi^0_{\rm II}l_{\rm II}\right)=\frac{\pi}{2}-\delta\;,
\label{eq:diffi} \\
\frac{1}{2}\left(\kappa_{\rm I}-\kappa_{\rm II}\right)
&=&\frac{1}{4}(\chi^{(3)}_{\rm I}l_{\rm I}
-\chi^{(3)}_{\rm II}l_{\rm II})=0\;.
\label{eq:difkappa}
\end{eqnarray}
\label{eq:dif}
\end{mathletters}
At the level of classical wave optics, when the operators are replaced
by {\it c\/}-numbers, these conditions ensure a fixed, $j$-independent
(intensity-independent) feedback factor for the $r$-mode, namely,
$R=\cos[(\theta_{\rm I}-\theta_{\rm II})/2]/ \sqrt{2} = \sin\delta/
\sqrt{2}$. By choosing $R\leq 0.1$ we cause the dynamics to be stable
[Fig.~\ref{fg:dyn}(b)] at the classical level. This result implies
that, classically, stabilization is achievable by additional rotation
and phase shift of the two-mode basis, forcing the system from the
unstable into the stable domain. These additional rotation and phase
shift are caused by an external ``device'' (the stabilizer) which,
although similar to the controlled system, is {\em not part of its
dynamics\/}, but rather the analog of the Ramsey apparatus which
rotates the states of a two-level system~\cite{kn:miwa}. If the
controlled system is near the edge of the stability domain
[$\cos(\theta_{\rm I}/2)$ just slightly larger than 0.14], then the
required additional rotation and phase shift $\theta_{\rm II}/2$ can
be small and amount to a {\em weak perturbation} of the parameters, as
in certain existing control methods~\cite{kn:ccw}. What singles out our
classical stabilization method is that it exploits the interference of
two correlated (coherent) evolutions.

The more intricate part of our proposal is concerned with correcting
for the {\em unavoidable} effects of quantum spread and entanglement,
which cause the breakdown of the quantum-classical correspondence in
phase space, when the Kerr effect is significant.  Upon inverting
Eq.~(\ref{eq:outin}) and using Eqs.~(\ref{eq:dif}), it can be readily
verified that the output states corresponding to the modes $r$ and $q$
will in general be entangled, because of the cross-terms
$\hat{a}^{\dag} \hat{b} + {\rm h.c.}$ and their equivalent
$\hat{f}^{\dag} \hat{e} + {\rm h.c.}$ in the phase-shift operators
$\hat{\theta}_{\rm I,II}$.

We are primarily interested in quasiclassical coherent-state inputs,
which can be readily compared to their classical counterparts
\begin{equation}
|{\rm in}\rangle=|\alpha\rangle_a|0\rangle_b=e^{-|\alpha|^2/2}
\sum_{n=0}^{\infty}\alpha^n\frac{(\hat{a}^{\dag})^n}{n!}
|0\rangle_a|0\rangle_b\;.
\label{eq:cohe}
\end{equation}
We then characterize our input density matrix by using the Q-function
quasiprobability distribution, which is diagonal in the coherent-state
basis, that is,
\begin{equation}
\hat{\rho}^{\rm (in)}_{F}=\int d^2\alpha d^2\beta P(\alpha,\beta)
|\alpha,\beta\rangle_{a,b}\langle\alpha,\beta|\;,
\label{eq:qpro}
\end{equation}
with $P(\alpha,\beta)\simeq g(\alpha-\alpha_{\rm in},\beta-0)$,
$g$ being a narrow Gaussian.
The resulting output, in general, will then be the entangled state
\begin{equation}
\hat{\rho}^{\rm (out)}_{F}=\int d^2\alpha d^2\beta P(\alpha,\beta)
|{\rm out}\rangle_{r,q}\langle{\rm out}|\;,
\label{eq:out}
\end{equation}
where the state $| \text{out} \rangle_{r,q}$ 
is in general an entangled state of field modes $r$ and $q$.

In order to achieve the same stabilization as in the classical limit,
our goal is to project out at each iteration a coherent state, and
feed it back from port $r$. It may be difficult to devise such a
measurement always, since the entangled state~(\ref{eq:out}) may have
a complicated phase-space distribution. We therefore choose to impose
an additional condition on the Kerr shift, corresponding to a {\em
fractional revival} of the Kerr shifted Fock states
$|n\rangle$~\cite{kn:rev}. The condition for a fractional revival is
$\kappa = \kappa_{\rm I} + \kappa_{\rm II} = \pi (m/l)$, where $m$ and
$l$ are {\em any\/} integers. Then, for sufficiently large $|\alpha|$
($|\alpha|\gg 1$), state~(\ref{eq:out}) corresponds to an entanglement
of a finite number of distinct coherent state components
(``Schr\"odinger kittens'') in modes $r$ and $q$ with different mean
phases and amplitudes. A phase-plane plot of such a state reveals the
origin of entanglement-induced chaotic behavior
[Fig.~\ref{fg:dyn}(b)-inset]: certain components of the fed-back
mixture may be in the chaotic domain. Such a state allows us to
project out the desired coherent-state component in $r$ (localized in
the regular domain), by a homodyne measurement of the $q$-component to
which it is correlated. We illustrate this process by imposing the
simplest (fractional) revival condition~\cite{kn:rev}
\begin{equation}
\kappa=\kappa_{\rm I}+\kappa_{\rm II}=\frac{\pi}{2} \;
\Longrightarrow \; e^{i\kappa n^2}=\left\{
\begin{array}{l} 1 \;\;\; n \; {\rm even} \\ i\;\;\; n\; {\rm odd}
\end{array} \right. \;.
\label{eq:cond}
\end{equation}
This condition is realizable for Kerr-shifts of photons interacting
with atoms in high-Q cavities \cite{kn:miwa}, and, in a more
straightforward fashion, for trapped ions interacting with lasers (see
below). Under this condition, we can show by a series of manipulations
that Eq.~(\ref{eq:out}) reduces to
\begin{eqnarray}
|{\rm out}\rangle_{r,q}&=&\frac{1}{\protect\sqrt{2}}
\left[e^{-i\pi/4}|\alpha e^{i\Phi}\cos\Phi\rangle_q
|-i\alpha e^{i\Phi}\sin\Phi\rangle_r\right.
\nonumber \\
 & & \;\;\; \left.+ e^{i\pi/4}|-i\alpha e^{i\Phi}\sin\Phi\rangle_q
 |\alpha e^{i\Phi}\cos\Phi\rangle_r \right]\;,
\label{eq:cat}
\end{eqnarray}
where $\Phi=(\varphi_{\rm I}+\varphi_{\rm II})/2$.  This entanglement
of two-mode Schr\"odinger-cat states implies that we have 50\%
probability of detecting the coherent state $|-i\alpha
e^{i\Phi}\sin\Phi\rangle_q$.  Such detection is known to be possible
by balanced homodyning, which mixes $|-i\alpha
e^{i\Phi}\sin\Phi\rangle_q$ with the appropriate local
oscillator~\cite{kn:miwa}. After a successful CM, we perform a
coherent mixing of the projected-out coherent state with the fixed
input via a 50\%-50\% beam splitter (Fig.~\ref{fg:scheme})
\begin{equation}
|\alpha e^{i\Phi}\cos\Phi\rangle_r + |\alpha_{\rm in} \rangle
\longrightarrow \left|\frac{\alpha_{\rm in}+\alpha
e^{i\Phi}\cos\Phi}{\protect\sqrt{2}}\right\rangle\;,
\label{eq:mix}
\end{equation}
finally feeding the resulting state [Eq.~(\ref{eq:mix})] back into the
$a$-port. This result reduces [upon using the phase-space distribution
(\ref{eq:out})] to its classical counterpart Eq.~(\ref{eq:itemap}) for
$|\alpha|\gg 1$.
The above analysis demonstrates the possibility of achieving complete
classical-like stabilization by an appropriate CM, thus overcoming the
effects of strong quantum entanglement. As stressed above, the price we
have to pay is that we have to perform as many trials as implied by the
success probability of the CM.

The outlined scheme can be fully implemented in an ion trap, where the
ion interacts with both traveling and standing-wave
fields~\cite{kn:ion}. A two-mode classical traveling-wave pulse of
area $\theta$ can perform an adiabatic rotation of two orthogonal
motional modes of the ion by the angle $\theta$: this is achievable
via coupling to two orthogonally polarized internal states, or by
alternative schemes, based on measurements of the internal
states~\cite{kn:ion}.
The rotation of the basis of two motional modes then mimics the effect
of beam splitters in the MZIs in Fig.~\ref{fg:scheme}. Kerr shift operators
of the motional states $\kappa \hat{a}^{\dag} \hat{a}$ are
emulated by an off-resonant standing-wave field, whose
$\kappa=\Omega^2\tau/\Delta$, $\Omega$ being its Rabi frequency, $\tau$ its
duration, and $\Delta$ the detuning. Schr\"odinger-cat motional states have
already been prepared in ion traps~\cite{kn:win}.

To conclude, we have presented a scheme which enforces regular
dynamical evolution towards a stable fixed point in an inherently
chaotic system by interference (phase correlations) with a
``stabilizer'' at the level of classical (wave) optics. At the level
of quantum description, the present scheme is the first attempt to
remedy the unwanted chaotic effects of quantum spread induced by
the system evolution and its entanglement with a ``stabilizer''
device. This is achieved by a conditional (homodyne) measurement of
the stabilizer output, which suppresses entanglement-induced chaos by
projecting out a quasiclassical coherent state localized in the
regular phase-space region. Conditions for such stabilization have
been presented for the quasiclassical analog of the Ikeda map
[Eqs.~(\ref{eq:dif}), (\ref{eq:cond})--(\ref{eq:mix})]. Although the
present scheme is quite distinct from existing schemes, it is in the
vein of those classical schemes of chaos control in which the phase
space is drastically changed, either by an external (driving) force
(the Pyragas method and its extensions) or by spectral filtering, for
the purpose of stabilization \cite{kn:ccs}.

We acknowledge the support of the German-Israeli Foundation and of the
TMR Network of the EU. M.~F. appreciates the warm hospitality and support
at the University of Ulm.

\end{document}